\documentstyle[wjh_bab,twocolumn]{article}
\title{Polarization Profiles of Scattered Emission Lines.
{\sc i}. General Formalism for Optically Thin Rayleigh Scattering}
\author{W. J. Henney\thanks{\here}}
\begin{document}


\def\ltae{\hbox{\raise.5ex\hbox{$<$}\kern-.8em\lower.5ex\hbox{$\sim$}$\,$}}
\def\gtae{\hbox{\raise.5ex\hbox{$>$}\kern-.8em\lower.5ex\hbox{$\sim$}$\,$}}
\def\abo{\sim\!\!}

\def\here{Dept.~of~Astronomy, U.~of~Manchester, M13~9PL}
\def\leeds{Dept.~of~Pure~and~Applied~Mathematics, U.~of~Leeds}
\def\mpropto{\!\propto\!}
\def\mgg{\!\gg\!}
\def\mll{\!\ll\!}
\def\meq{\!=\!}
\def\msim{\!\sim\!}
\def\mplus{\!+\!}
\def\mprop{\!\propto\!}
\def\msimeq{\!\simeq\!}
\def\mtimes{\!\times\!}
\def\mleq{\!\leq\!}
\def\mlt{\!<\!}
\def\mdiv{\!\/\!}
\def\mmin{\!-\!}
\def\({\left(}
\def\){\right)}
\def\sq{\!^{2}}
\def\erf{\mbox{erf}}
\def\etal{{\rm et~al.}\ }
\def\ie{{\rm i.e.}\ }
\def\eg{{\rm e.g.}\ }
\def\etc{{\rm etc.}\ }
\def\kms{{\rm\,km\,s^{-1}}}
\def\pcc{{\rm\,cm^{-3}}}
\def\ergps{{\rm\,erg\,s^{-1}}}
\def\persec{{\rm\,s^{-1}}}
\def\fu{{\rm\,erg\,\,cm^{-2}s^{-1}}}
\def\cms{\rm\,cm\,s^{-1}}
\def\kms{\rm\,km\,s^{-1}}
\def\gcm{{\rm g\,cm^{-3}}}
\def\mic{\,\mu\rm m}
\def\cm{\,\rm cm}
\def\yr{\,\rm yr}
\def\cmsq{\,\rm cm^2}
\def\pcmsq{\,\rm cm^{-2}}
\def\AU{{\footnotesize\,AU}}
\def\au{{\footnotesize\,AU}}
\def\mdot{\dot{M}_{\star}}
\def\mw{\dot{M}_{_{\rm W}}}
\def\r0{\rho_{_{0}}}
\def\lw{L_{_{\rm W}}}
\def\vw{V_{_{\rm W}}}
\def\vstar{V_{\star}}
\def\smy{\rm\,M_{\odot}\,yr^{-1}}
\def\sm{\rm\,M_{\odot}}
\def\rs{R_{_{\rm S}}}
\def\ls{\Lambda_{_{\rm S}}}
\def\ra{r_{_{A}}}
\def\mmira{\dot{M}_{_{Mira}}}
\def\vmw{v_{_{\rm W}}}
\def\rhow{\rho_{_{\rm W}}}
\def\vrel{v_{_{\rm rel}}}
\def\macc{\dot{M}_{\rm acc}}
\def\us{{\cal U}_{_{\rm S}}}
\def\le{\Lambda_{_{\rm e}}}
\def\ue{{\cal U}_{_{\rm e}}}
\def\lp{\Lambda_{_{\rm p}}}
\def\up{{\cal U}_{_{\rm p}}}
\def\seca{\(1\mplus\(\frac{1}{\ls}\frac{\partial\ls}{\partial\theta}\)^{2}\)
^{\frac{1}{2}}}
\def\sech{{\rm sech}}
\def\ms{M_{_{\rm S}}}
\def\vs{V_{_{\rm S}}}
\def\mh{{\rm m_{_{\rm H}}}}
\def\thi{{\theta}_{i}}
\def\yi{{\cal Y}_{i}}
\def\ui{{\cal U}_{i}}
\def\dd{{\rm d}}
\def\rc{R_{\rm c}}
\def\zc{z_{\rm c}}
\def\bx{\bar{x}}
\def\by{\bar{y}}
\def\bz{\bar{z}}
\def\nh{n_{\rm H}}
\def\sfrac#1#2{{#1}/{#2}}
\newcommand{\bvec}[1]{\mbox{\protect\boldmath $#1$}}
\newcommand{\uvec}[1]{{\bf\hat{\bvec{#1}}}}
\def\vecu{\mbox{\protect\boldmath $\displaystyle u$}_0}
\def\zhat{\hat{\mbox{\boldmath $z$}}}
\def\rhat{\hat{\mbox{\boldmath $R$}}}
\def\strut{\rule{0pt}{24pt}}
\def\halfstrut{\rule{0pt}{12pt}}
\def\smallstrut{\rule{0pt}{6pt}}
\def\lift{\rule[-6pt]{0pt}{18pt}}
\def\dprod{\mbox{}_\bullet}
\def\bzc{\bz_{\rm c}}
\def\vt{\tilde{v}}
\def\sigmat{\mbox{\large\boldmath$\sigma$}_{\Omega}}
\def\csqbt{\cos^{2}\Theta}
\def\ifac{\frac{\tau I_0}{4\pi}}
\def\thc{\theta_{\rm c}}
\def\half{\frac{1}{2}}
\def\fwhm{\Delta v_{\scriptscriptstyle \rm FWHM}}
\def\\mn {Mon. Not. R. ast. Soc.}
\def\\apj {Astrophys. J.}
\def\\aa {Astr. Astrophys.}
\def\\aj {Astr. J.}
\def\\pasp {Publs. astr. Soc. Pacif.}
\def\textfraction{0.05}
\def\topfraction{0.99}
\def\bottomfraction{0.99}
\def\floatsep{0.25in}
\def\textfloatsep{0.25in}

\def\eg{e.\ g.}
\def\rad{{\rm rad}}
\def\etal{at~al.\ }
\def\papi{Paper~I}
\def\papii{Paper~II}
\def\papiii{Paper~III}
\def\papiv{Paper~IV}
\def\NRAL{Nuffield Radio Astronomy Laboratories, University of
Manchester, Jodrell Bank, Macclesfield, Cheshire, SK11 9DL, UK}
\def\here{Instituto de Astronom\'{\i}a, UNAM, Apartado Postal 70--264,
04510 M\'{e}xico D. F., M\'{e}xico}
\def\UMIST{Department of Mathematics, UMIST, PO Box 88, Manchester, M60~1QD,
UK}
\def\STScI{Space Telescope Science Institute, 3700 San Martin Drive,
Baltimore, MD 21218, USA}
\bibliographystyle{apj}
\begin{titlepage}
\begin{centering}
{\LARGE \sc Polarization Profiles of Scattered}\\ \vspace*{5pt}
{\LARGE \sc  Emission Lines.}\\ \vspace*{5pt}
{\LARGE \sc {I}. General Formalism for}\\ \vspace*{5pt}
{\LARGE \sc  Optically Thin}\\ \vspace*{5pt}
{\LARGE \sc  Rayleigh Scattering}\\
\vspace*{26mm}
{W. J. Henney}\\
\parbox[t]{10cm}{\centering\here} \\
\vspace*{9cm}
{\em To be published in The Astrophysical Journal, May 20 1994}\\
\end{centering}
\end{titlepage}
\begin{centering}
{\Large \sc Abstract}\\
\vspace*{8mm}
\end{centering}
{
A general theoretical framework is developed for interpreting
spectropolarimetric observations of optically thin emission line scattering
from small dust particles.
Spatially integrated and spatially resolved line profiles
of both scattered intensity and polarization are calculated
analytically from a variety of simple kinematic models. These
calculations will provide a foundation for further studies of emission
line scattering from dust and electrons in such diverse astrophysical
environments as Herbig-Haro objects, symbiotic stars, starburst
galaxies and active galactic nuclei.
}

\noindent {\em Subject headings}: ISM: dust,
extinction---ISM: jets and outflows---ISM: reflection
nebulae---polarization

\section{Introduction}
The Doppler shifts of
optical emission lines which have been scattered by surrounding dust
and electrons
can provide useful information about the kinematics, geometry, and physical
conditions of astrophysical flows. In
principle, the scatterers can provide views of the line-emitting gas
from different directions, allowing the 3-dimensional velocity of the
emitting gas to be determined and revealing sources which are obscured
from direct view. Unfortunately, the interpretation of these scattered
emission lines is not straightforward since, in general,
the geometry of the scattering and the velocity of the scatterers is
unknown. If spectropolarimetric observations are available, then
determination of the relative orientation and velocity of the source
and the scatterers becomes more plausible, since the scattered light will be
partially polarized to a degree dependent on the angle of scattering
and on the details of the scattering process.

In this paper, a variety of very simple models of the scattering geometry
and kinematics are investigated. Polarimetric line profiles are
calculated (both spatially integrated and spatially resolved) for the
cases of Rayleigh scattering by small dust particles, where the
scattering phase function has a particularly simple form and it is
feasible to derive analytic expressions for the scattered line
profiles.

Previous work on the problem of the
scattering of emission lines by dust or electrons has tended to
involve detailed numerical radiative transfer modelling, often
using Monte Carlo codes. Examples of the use of such techniques can be
found applied to, \eg$\!$, electron scattering in Wolf-Rayet winds
\cite{Hil91} or scattering in circumstellar dust shells
\cite{Lef92,Mea92}. While such approaches have obvious merits, they
are limited in their applicability, often needing to be specially
tailored for each object, and it is difficult to draw general
conclusions from them. A simpler approach is taken by \scite{Woo93}, who
model electron scattering in the circumstellar disks of Be stars, but,
even here, numerical techniques are employed. In the present paper,
numerical calculations are eschewed for the most part and scattered
line profiles of intensity and polarization are calculated
analytically. The price that must be paid in order to make this
approach tractable is to restrict one's attention to optically thin
Rayleigh scattering of narrow emission lines which originate from a
point source. The chief advantage of this approach is that it makes it
possible to investigate a wide  range of models for the geometry and
kinematics of the scatterers and to develop a
broad understanding of the ways in which these affect the line
profiles.
This understanding can then provide a firm foundation for further
investigation into the effects of different phase functions (dust),
thermal broadening (electrons), an extended source, absorption,
multiple scattering and other complications. This development will be
carried out in further papers of this series and detailed applications
of the resultant techniques will be made to particular cases such as
upstream dust scattering in Herbig-Haro objects \cite{Nor91}, dust scattering
in the jet of the symbiotic star R Aquarii \cite{Sol92}, dust
scattering  in the
superwinds of starburst galaxies \cite{Sca91} and electron scattering
in active galactic nuclei \cite{Mil91}. Some preliminary results in
these areas can be found in \scite{th:Hen92}.

\section{Rayleigh Scattering}
\label{phase}
Arbitrarily polarized light is fully described by means of the
four-component
Stokes vector \cite{Sto52} ${\bf I} \meq [I,Q,U,V]^T$, which can be
functionally
defined in terms of measurements of the light intensity as follows: $I$
is the total intensity of the light; $Q$ is the difference between the
intensities after the light has passed through a linear dichroic plate
with its transmitting axis aligned respectively perpendicular and
parallel to a reference direction; $U$ is the difference between
the intensities with the same plate aligned at $\pm45^{\circ}$ to the
reference direction, and $V$ is the difference between the intensities
after the light has passed through a right and left handed circular
dichroic plate respectively.
Unpolarized light has $Q\meq U\meq V \meq 0$ and fully polarized light
has $I^2\meq Q^2 \mplus U^2 \mplus V^2$. Partially polarized light has
a degree of polarization
\begin{equation}
p \meq (Q^2+U^2+V^2)^{1/2}{\big /}I
\label{pdef}
\end{equation}
and an angle of polarization $\chi$ (with respect to the reference
direction in the definition of $Q$ and $U$ above) given by
\begin{equation}
\tan 2 \chi \meq U/Q\strut \label{chidef2}.
\end{equation}
Note that, since one must choose a reference direction corresponding
to $\chi\meq0$, there is a certain arbitrariness in the specification
of $Q$ and $U$. In astronomy this reference direction is conventionally
taken to be North (or sometimes, the galactic plane).
 The effect on the Stokes vector of a rotation of the
reference direction by an angle $\phi$ is to multiply it by the
rotation matrix
\begin{equation}
{\bf M}(\phi) \meq
\left[ \begin{array}{cccc}
       1& 0 & 0 & 0\\ 0 & \cos 2 \phi& \sin 2 \phi & 0 \\ 0 & -\sin 2
\phi & \cos 2 \phi& 0 \\ 0 & 0 & 0 & 1 \end{array} \right].
\end{equation}

In the following discussion, it is imagined that
light is emitted from a source at the origin
and scattered by a single scatterer at $[x,y,z]$ toward an observer
who is in the far field and is looking down the negative $z$-axis (see
Figure~\ref{scat1}).
\begin{figure}
  \begin{center}
    \makebox[4.722in][l]{
  \vbox to 3.597in{
    \vfill
    \includegraphics{fig01.ps}
  }
  \vspace{-\baselineskip}
}

    \caption{\sf Geometry of single scattering. Light travels from a
      source at the origin ({\sc o}) in a direction $z^{\prime}$ until
      it is scattered at a point {\sc b} into a new direction $z$,
      towards the observer.  Perpendicular axes $x$ and $y$ are in the
      plane of the sky such that $yz$ is the observer's meridional
      plane.  The perpendicular projection of {\sc b} onto the
      $z$-axis is {\sc a} and onto the $xy$-plane is {\sc c}. The
      rectangle {\sc oabc} thus lies in the scattering plane,
      containing the incident and scattered rays. The scattering angle
      between the two rays is $\Theta$ and the position angle of the
      scattering plane is $\psi$. $x^{\prime}$ and $y^{\prime}$
      complete an orthogonal set of axes with $z^{\prime}$ such that
      $y^{\prime}z^{\prime}$ is the meridional plane for the incident
      light.  The perpendicular projection of {\sc c} onto the
      $x^{\prime}y^{\prime}$-plane is then {\sc d}, which is coplanar
      with {\sc oabc}, so that the position angle of the scattering
      plane in the primed coordinate system is $\psi^{\prime}$.}
    \label{scat1}
  \end{center}
\end{figure}
When the light is scattered, the components of the scattered Stokes
vector are linearly related to those of the incident Stokes vector. Hence,
the scattering cross-section can be written as a matrix.
One way of representing the differential cross-section per solid angle
$\Omega $ is then
\begin{equation}
\sigmat \meq \frac{3\sigma_0}{16\pi} {\bf K},
\end{equation}
where $\sigma_0$ is the mean scattering cross-section and the matrix
${\bf K}$ (the scattering kernel; \pcite{b:Pom73}) contains the
direction-dependent scattering information.
The matrix $\sigmat$ is normalized so that
\begin{equation}
\int_{4\pi} \sigmat {\bf I} \dd\Omega \meq \sigma_0 {\bf I}
\label{4pi}
\end{equation}
for any Stokes vector ${\bf I}$.
 If ${\bf I}_0$ is the
Stokes vector of the unscattered light at the source (in the direction
of the scatterer), then, by considering the solid angle at the source
subtended by the scatterer, the scattered Stokes vector will be
\begin{equation}
{\bf I}_{\rm S} \meq \sigmat{\bf I}_0/R^2,
\end{equation}
where $R^2\meq x^2
\mplus y^2 \mplus z^2$.
If the scatterers are much smaller than the wavelength of the light,
then, {\em so long as both are
specified with respect to the scattering plane}, the Stokes vector of
the scattered light is related to that of the incident light via the
Rayleigh scattering matrix \cite[p~37]{b:Cha60}:
\begin{equation}
{\bf R}(\Theta) \meq \half
\left[ \begin{array}{cccc}
       \csqbt \mplus 1&  \csqbt \!-\! 1 & 0 & 0 \\
\csqbt \mmin 1& \csqbt \mplus 1& 0 & 0  \\ 0 & 0 &
2\cos \Theta & 0 \\ 0 & 0 & 0 & 2\cos \Theta \end{array} \right]
\label{rayleigh}
\end{equation}
where $\Theta$ is the scattering angle, the angle between the incident
and the scattered ray.  In general, if the scattering plane is at an
angle $\psi$ to the reference direction, then ${\bf R}$ must be pre-
and post-multiplied by appropriate rotation matrices ${\bf M}$ to give
the scattering kernel ${\bf K}(\Theta,\psi)$.  In our example, the
scattering kernel is
\begin{equation}
{\bf K}(\Theta,\psi) \meq {\bf M}(\pi-\psi) {\bf R}(\Theta) {\bf
M}(-\psi\prime)
\end{equation}
where the angles $\psi$ and $\psi\prime$ are shown in
Figure~\ref{scat1} and the reference direction is taken to be the
$y$-axis.
If the incident light is unpolarized, then $Q$, $U$, and $V$ are zero
in the incident ray, so the ${\bf M}(-\psi\prime)$ rotation need not
be carried out and the Stokes vector of the scattered light is given
by
\begin{equation}
\label{iscat1}
{\bf I}_{\rm S} \meq\frac{\sigma_0 I_0 }{4\pi R^2} {\bf X}(\Theta,\psi),
\end{equation}
where
\begin{equation}
 {\bf X}(\Theta,\psi) \meq \frac{3}{4} \left[ \begin{array}{c}
\cos^2\Theta \mplus 1\\
-\cos 2 \psi \sin^2\Theta\\
\sin 2 \psi \sin^2\Theta\\
0 \end{array}\right].
\label{ray_phase}
\end{equation}
 The vector scattering phase function ${\bf X}(\Theta,\psi)$ is
normalized so that, when integrated over all
solid angles, the $I$-component becomes $4\pi$ and all other components
become zero. Since single scattering of initially unpolarized light
cannot produce any $V$-component to the scattered Stokes vector, the
$V$-component will be dropped in the rest of the paper and a reduced
Stokes vector ${\bf I} \meq [I,Q,U]^T$ will be used.
{}From equations~(\ref{ray_phase}) and~(\ref{pdef}) it can be seen that the
degree of polarization of Rayleigh scattered light is
\begin{equation}
p\meq \sin\sq\Theta/(\cos\sq\Theta\mplus 1)
\label{pray}
\end{equation}
while the angle of polarization $\chi$ is, using equation~(\ref{chidef2}),
\begin{equation}
\chi \meq \phi \pm \pi/2.
\label{chiray}
\end{equation}
Hence the scattered light is always partially polarized {\em
perpendicular}
to the plane of scattering and is fully polarized for scattering
angles of 90$^{\circ}$.
\section{Simple Scattering Geometries}
\label{geom}
In this section, three simple classes of problems are considered as
examples of Rayleigh scattering: first, that of a stationary dust
cloud scattering light from a moving unpolarized source at its center;
second, that of an outflowing dusty wind; and third, that of a
free-falling dusty inflow, the last two both scattering light from a
stationary unpolarized central source.  In the first instance, the
scattering cloud/wind/infall is assumed to be spherically symmetric
and the light source is assumed to be a point, although the first of
these assumptions is subsequently relaxed for the first two cases.

Considering the spatially integrated light first, the three models are
equivalent and it is evident on
symmetry grounds that the scattered light  will have no net polarization.
This can
be verified by taking the scattered intensity from a single scatterer
(eq.~[\ref{iscat1}]) and integrating it over all the scatterers in
the cloud. If the number density of scatterers is $n(R)$ and the cloud
radius is $\rc$ then this yields
\begin{equation}
{\bf I}_{\rm S}\meq \frac{3 \sigma_0 I_0 }{16 \pi}
\int_{\mbox{cloud}} \frac{n(R)}{R^2}{\bf X}(\Theta,\psi)
\dd V
\meq \tau I_0
\left[ \begin{array}{c} 1\\0\\0 \end{array} \right]
\meq \tau {\bf I}_0
\label{totint}
\end{equation}
where $\tau\meq \sigma_0 \int_0^{\rc} n\, \dd R$ is the scattering
optical depth  to the source.
Of course, since $\tau\ll 1$, the scattered light makes a negligible
contribution to the total intensity unless the direct light from the
source is obscured.

\subsection{Spatially Integrated Scattered Line Profiles}
\label{velscat}
\subsubsection{Moving Source in a Stationary Scattering Cloud}
\label{mover}
If it is now supposed that the source of light is moving with a
velocity $\vecu$ with respect to the (stationary) dust cloud and that
the light it emits is monochromatic (of frequency $\nu_0$), then the
scattered light will be Doppler shifted by $\Delta\nu\meq(\nu_0/c)
\rhat \cdot \vecu$, where $\rhat$ is the unit vector in the direction
of the scatterer from the source.  One can define a dimensionless
velocity shift\footnote{\parbox[t]{8cm}{\
In keeping with established convention, red shifts
will be treated as positive.}}
(frequency shift) of the scattered light as $v\meq-\rhat
\cdot \vecu / u_0 \equiv -\cos \gamma$. This is illustrated in
\begin{figure}
  \begin{center}
    \makebox[4.722in][l]{
  \vbox to 4.847in{
    \vfill
    \includegraphics{fig02.ps}
  }
  \vspace{-\baselineskip}
}

    \caption{\sf Geometry of scattering from a moving source. The source
      velocity $\vecu$ lies in the $xz$-plane and is inclined at an
      angle $\alpha$ to the plane of the sky ($xy$). The radius from
      the source to the scatterer makes an angle $\gamma$ with
      $\vecu$. The azimuth of the scatterer, measured about $\vecu$
      from a direction parallel to the $y$-axis, is $\xi$. The dotted
      lines show the ``isovelocity cone'', on which all scatterers see
      the same projected source velocity $u_0\cos\gamma$. }
    \label{vel_diag}
  \end{center}
\end{figure}
Figure~\ref{vel_diag}, in which the $x$-axis is taken to be along the
projection of $\vecu$ onto the plane of the sky and the angle between
$\vecu$ and the $x$-axis is denoted by $\alpha$.
It is apparent that all scatterers that lie on the surface of a cone
of opening half-angle $\gamma$ ``see'' the same Doppler shift
($v\meq-\cos\gamma$) in the light from the source. This will be called
the ``isovelocity cone'' and $\xi$ will signify the angle around this
cone, measured from a direction parallel to the $y$-axis (see
Figure~\ref{vel_diag}).

The spatially integrated scattered line shape can best be determined
by considering the relationship between the two angular co-ordinate
frames $[\gamma,\xi]$ and $[\Theta,\psi]$. Equating Cartesian
components of a unit vector in the two co-ordinate frames gives
\begin{subeqnarray}
x \meq  \sin\Theta\sin\psi & \meq & \sin\gamma\sin\xi\sin\alpha
\mplus \cos\gamma\cos\alpha \\
y \meq \sin\Theta\cos\psi & \meq & \sin\gamma\cos\xi \\ z  \meq
\cos\Theta & \meq & \sin\gamma\sin\xi\cos\alpha \mmin
\cos\gamma\sin\alpha
\end{subeqnarray}
so that, using equation~(\ref{ray_phase}), the scattering phase function
becomes
\begin{equation}
{\bf X}(\gamma,\xi)\meq \frac{3}{4}
\left[ \begin{array}{c} a_0 \mplus a_1 \sin\xi
\mplus a_2 \sin\sq\xi \\ b_0 \mplus b_1 \sin\xi \mplus b_2 \sin\sq\xi
\mplus b_3 \cos\sq\xi\\ c_1 \sin\xi \mplus c_4 \sin\xi\cos\xi
\end{array} \right]
\end{equation}
where{\small
\begin{displaymath}
\begin{array}{lll}
a_0 \meq\sin\sq\alpha\cos\sq\gamma \mplus 1 ;\!\!& a_1\meq -\half \sin
2\alpha \sin 2\gamma;\!\!& a_2\meq \cos\sq\alpha\sin\sq\gamma;\!\!\\ b_0\meq
\cos\sq\alpha\cos\sq\gamma;\!\!& b_1\meq \half\sin2\alpha\sin2\gamma;\!\!& b_2
\meq\sin\sq\alpha\sin\sq\gamma ;\!\!\\  b_3 \meq -\sin\sq\gamma;\!\! & c_1\meq
-\cos\alpha\sin2\gamma;\!\!& c_4 \meq -2 \sin\alpha\sin\sq\gamma.
\end{array}
\end{displaymath}
}
To find the scattered intensity in a velocity range $v_1$ to $v_2$ it
is necessary to integrate equation~(\ref{iscat1}) over all the dust
between the two isovelocity cones corresponding to $v_1$ and $v_2$.
Hence,
\begin{equation}
{\bf I}(v_1 \rightarrow v_2) \meq \ifac \int_{\gamma =
\pi\mmin\cos^{-1}\!v_1}^{\pi\mmin\cos^{-1}\!v_2}
\int_{\xi = 0}^{2\pi} {\bf X}(\gamma,\xi) \sin\gamma\, \dd\gamma\, \dd\xi.
\end{equation}
Since ${\dd} v \meq \sin\gamma \,\dd\gamma$, the specific intensity
(per unit velocity range) is then
\begin{equation}
{\bf I}_v \meq \left| \frac{\dd {\bf I}\,}{\dd{\mit v}} \right| \meq \ifac
\int_0^{2\pi} {\bf X}(v,\xi)\, \dd \xi
\meq \frac{3}{8}\tau I_0 \left[ \begin{array}{c}
                              a_0\mplus \half a_2 \\ \!\!\!b_0\mplus \half
b_2 \mplus \half b_3\!\!\!\\ 0\end{array}\right]
\label{specintens}
\end{equation}
so that
\begin{subeqnarray}
\slabel{integvel}
I_v &\meq& \frac{3}{16}\tau I_0 \left\{ 2\(1\mplus v^2\)
\mplus \(1-3v^2\)\cos\sq\alpha\right\}\strut,\\
\slabel{integvel_q}
Q_v &\meq& \frac{3}{16}\tau I_0 \(1-3v^2\)\cos\sq\alpha\strut,\\
U_v &\meq &0 \strut.
\end{subeqnarray}
  The polarization of the scattered light is
then
\begin{equation}
\label{integvel_p}
p_v\meq \frac{1-3v^2 \cos\sq\alpha}{2\(1\mplus
v^2\)\mplus\(1-3v^2\)\cos\sq\alpha},
\end{equation}
while the angle of polarization $\chi$ is zero (perpendicular to the
projected source velocity) when $Q_v$ is positive, or 90$^\circ$
(parallel to the projected source velocity) when $Q_v$ is
negative.\footnote{
\parbox[t]{8cm}{\ Notice that the
polarization is here allowed to take the sign of $Q$, rather than
being constrained to be positive. This is a useful convention in
instances like this one, where $\chi$ can only take one of two values,
but can be distracting otherwise.}}
These quantities are plotted in
\begin{figure}
  \begin{center}
    \makebox[7.305in][l]{
  \vbox to 2.375in{
    \vfill
    \includegraphics{fig03.ps}
  }
  \vspace{-\baselineskip}
}

    \caption{\sf Line profiles of spatially integrated light, Rayleigh
      scattered by a stationary spherical dust cloud that surrounds a
      moving source, for various inclinations $\alpha$ of the source
      velocity to the plane of the sky. Solid line is specific
      intensity and dashed line is degree of polarization.}
    \label{intspec}
  \end{center}
\end{figure}
Figure~\ref{intspec} for various values of $\alpha$.

It should be noted that equation~(\ref{integvel_p}) is more general than
equations~(\ref{integvel}) and~(\ref{integvel_q}) and applies to any
distribution of scatterers that has cylindrical symmetry about the
direction of the source velocity. This is because\label{general} the
number density $n(R,\gamma,\xi)$ cancels out when taking the ratio of
$Q$ and $I$. (Cylindrical symmetry is still required, since it is assumed
when performing the $\xi$ integral
in eq.~[\ref{specintens}].)

To illustrate how the intensity profile is modified if the scattering
cloud is not spherical, the case of prolate and oblate ellipsoidal
clouds will be considered.
\paragraph{Ellipsoidal Scattering Clouds:}
If the scattering cloud is non-spherical, then the chance of a source
photon being scattered is dependent on its initial direction.
For a prolate ellipsoid of eccentricity $e$, with symmetry axis aligned with
the source velocity, this can be translated into a direction dependent
optical depth of
\begin{equation}
\frac{\tau(\gamma)}{\tau_\star}\meq\(\frac{1-e^2\sin^2\alpha}
{1-e^2 \cos^2\gamma}\)^{1/2},
\end{equation}
where $\tau_\star$ is the optical depth along the line of sight from
the observer (corresponding to $\gamma\meq\pi/2
\mmin\alpha$).
 Hence, the intensity of the scattered light is given by
\begin{eqnarray}
I_v & \meq & \frac{3}{8}\tau_\star I_0
\(\frac{1-e^2\sin^2\alpha}{1-e^2 v^2}\)^{1/2}\nonumber \\
& & \times
 \left\{ 2\(1\mplus v^2\)
\mplus \(1-3v^2\)\cos\sq\alpha\right\}.
\end{eqnarray}
A similar treatment for an oblate ellipsoid yields
\begin{eqnarray}
I_v &\meq &\frac{3}{8}\tau_\star I_0
\(\frac{1-e^2\sin^2\alpha}{\(1-e^2\)\mplus e^2 v^2}\)^{1/2}\nonumber \\
& & \times
 \left\{ 2\(1\mplus v^2\)
\mplus \(1-3v^2\)\cos\sq\alpha\right\}.
\end{eqnarray}
The resultant line shapes are plotted for various values of $e$ and
\begin{figure}
  \begin{center}
    \makebox[7.847in][l]{
  \vbox to 2.402in{
    \vfill
    \includegraphics{fig04.ps}
  }
  \vspace{-\baselineskip}
}

    \caption{\sf Same as Figure~\protect\ref{intspec} but for prolate
spheroidal
      scattering clouds.  Different line types correspond to different
      eccentricities, as indicated in the key.}
    \label{prol_vel}
  \end{center}
\end{figure}
\begin{figure}
  \begin{center}
    \makebox[7.847in][l]{
  \vbox to 2.402in{
    \vfill
    \includegraphics{fig05.ps}
  }
  \vspace{-\baselineskip}
}

    \caption{\sf Same as Figure~\protect\ref{prol_vel} but for oblate
      spheroidal scattering clouds.}
    \label{ob_vel}
  \end{center}
\end{figure}
$\alpha$ in Figures~\ref{prol_vel} and~\ref{ob_vel}. The polarization
is not shown since it is the same as for a spherical cloud.  It is apparent
that prolate ellipsoids produce line shapes in which most of the
intensity is concentrated toward $v\meq\pm1$ whereas oblate ellipsoids
produce line shapes strongly peaked at $v\meq0$. The reason for this
can be seen if one considers the distributions of scatterers in the two
cases: in prolate clouds, most of the scatterers are concentrated
along the direction of the source velocity, either ``upstream'' or
``downstream'', while in oblate clouds they are concentrated toward
the plane perpendicular to the source velocity and hence see little
Doppler shift. It will be noticed that the integrated intensity (with
respect to $v$) is not the same in each case.  This is because the
intensities are normalized with respect to $\tau_\star I_0$, where
$\tau_\star$ is the scattering optical depth to the source along the
line of sight. Generally, $\tau_\star$ is not equal to the
angle-averaged optical
depth (weighted by the phase function) $\bar{\tau}$, although it is this latter
that determines the integrated scattered intensity.  For example, a
prolate ellipsoid at $\alpha\meq0$ or an oblate ellipsoid at
$\alpha\meq\pi/2$ would each have $\tau_\star < \bar{\tau}$.

\subsubsection{Constant Velocity Scattering Wind}
\label{convel1}
If the scattering wind is supposed to flow radially outward at
constant speed $u_{\rm w}$, then the Doppler frequency shift in the
scattered light will be $\Delta\nu\meq(\nu_0/c)u_{\rm w}(1\mplus \rhat
\cdot \zhat)$, where $\zhat$ is the unit vector along the $z$-axis
(from the source toward the observer). In the same manner as for the
moving source model, a dimensionless velocity shift $v\meq 1\mplus \rhat
\cdot \zhat $ is introduced which again defines isovelocity cones. In
this model, however, the opening half-angle of the cone is the
scattering angle $\Theta$ and this is related to the velocity shift by
$\cos\Theta\meq v-1$.
Determination of the scattered line profile is hence far simpler than
in the preceding section, giving
\begin{equation}
{\bf I}_v \meq \ifac
\int_0^{2\pi} {\bf X}(v,\Psi)\, \dd \Psi
\meq \frac{3}{8}\tau I_0 \left[ \begin{array}{c}
                              v^2-2v\mplus 2\\ 0\\ 0 \end{array}\right].
\label{specintens2}
\end{equation}
The scattered line is unpolarized because the isovelocity cones
project onto the plane of the sky as circles. This line shape is illustrated in
\begin{figure}
  \begin{center}
    \makebox[2.333in][l]{
  \vbox to 2.125in{
    \vfill
    \includegraphics{fig06.ps}
  }
  \vspace{-\baselineskip}
}

    \caption{\sf Spatially integrated line profile from a
      constant-velocity spherically symmetric wind of scatterers.}
    \label{sp_wind_line}
  \end{center}
\end{figure}
Figure~\ref{sp_wind_line} and it can be seen that it is identical to
the moving source model with $\alpha\meq \pi/2$, except for a shift in
the velocity origin.
If the wind is not spherically symmetric but is conical or disk-like
in form, then the symmetry of the isovelocity cones is broken and
polarized line profiles can result.
\paragraph{Outflowing Bicone:}
The outflow axis of the wind is taken to make an angle $\alpha$ with
the plane of the sky and the wind density is assumed to be
independent of direction within the two cones of opening half-angle
$\delta_{\rm c}$ about this axis, and zero outside the cones.
 These scattering cones will
intersect a given isovelocity cone along 0, 2 or 4 radii with position
angles ($\Psi_1 \dots \Psi_4$) given by
\begin{equation}\label{psicone}\begin{array}{ccccc}
\strut\sin\Psi_1&\meq&
\sin\Psi_2&\meq&\frac{\displaystyle\rule[-4pt]{0pt}{14pt}
\cos\delta_{\rm c}-\cos\theta\sin\alpha}{\displaystyle\rule[-4pt]{0pt}{14pt}
\sin\theta \cos\alpha},\\
\strut\sin\Psi_3&\meq&
\sin\Psi_4&\meq&\frac{\displaystyle\rule[-4pt]{0pt}{14pt}
-\cos\delta_{\rm c}-\cos\theta\sin\alpha}{\displaystyle\rule[-4pt]{0pt}{14pt}
\sin\theta\cos\alpha},\\
\end{array}
\end{equation}
where $\theta\meq\cos^{-1}|v-1|$.
The scattered Stokes vector is therefore given by
\begin{equation}
{\bf I}_v \meq \ifac
\left\{ \int_{\Psi_1}^{\Psi_2} {\bf X}(v,\Psi)\, \dd \Psi \mplus
\int_{\Psi_3}^{\Psi_4} {\bf X}(v,\Psi)\, \dd \Psi \right\},
\end{equation}
which can be integrated to give the following:
\begin{subeqnarray}
\slabel{integvel_icone}
I_v &\meq& \frac{3}{16\pi}\tau I_0(v^2\mmin 2v\mplus 1)
F_1\rule[-3ex]{0ex}{6ex}\\
\slabel{integvel_qcone}
Q_v &\meq& \frac{3}{16\pi}\tau I_0 (2v\mmin v^2) F_2\rule[-3ex]{0ex}{6ex}\\
U_v&\meq&0\rule[-3ex]{0ex}{6ex},
\end{subeqnarray}
where
\begin{subeqnarray}
  F_1 \meq \left\{
  \begin{array}{ll}
    2\pi& \left[\mbox{\bf if} \ \ \ \alpha\mmin\theta\geq \pi/2 \mmin
    \delta \right]\lift \\
    2\pi\mmin 2(\Psi_1\mmin\Psi_3)& \left[
    \mbox{\bf if} \ \ \ \theta\mmin\alpha\geq \pi/2 \mmin \delta \right]\lift
\\
    0 & \begin{minipage}{3.5cm}$\left\{\begin{array}{l}\left[\mbox{\bf if} \
    \ \ \alpha\mplus\theta > \pi/2 \mplus
    \delta \right. \\ \left. \ \mbox{\bf or} \ \ \ \alpha\mplus\theta <
    \pi/2 \mmin \delta \right]\end{array}\right. $ \end{minipage}  \\
    \pi\mmin 2\Psi_1 & \left[\mbox{\bf otherwise}\right]\lift
  \end{array} \lift\right. \\
  \strut\hfill{  } \nonumber \\
  F_2 \meq \left\{
  \begin{array}{ll}
    0 &
    \left[ \mbox{\bf if} \ \ \ \alpha\mmin\theta\geq \pi/2
    \mmin\delta\right]
    \lift \\ \sin 2\Psi_1\mmin\sin 2\Psi_3&\left[
    \mbox{\bf if} \ \ \ \theta\mmin\alpha\geq \pi/2 \mmin
    \delta\right]
    \lift\\ 0 &
    \begin{minipage}{3.5cm}$\left\{\begin{array}{l}\left[\mbox{\bf if} \ \
      \ \alpha\mplus\theta > \pi/2 \mplus \delta \right. \\ \left. \
      \mbox{\bf or} \ \ \ \alpha\mplus\theta < \pi/2 \mmin \delta
    \right]\end{array}\right. $ \end{minipage}\\
    \sin 2\Psi_1 & \left[\mbox{\bf otherwise}\right]\lift
  \end{array} \lift\right.
\end{subeqnarray}
\begin{figure}
  \begin{center}
    \makebox[5.222in][l]{
  \vbox to 4.750in{
    \vfill
    \includegraphics{fig07.ps}
  }
  \vspace{-\baselineskip}
}

    \caption{\sf Integrated spectra of intensity (solid line) and degree of
      polarization (dashed line) for a scattering biconical wind with cone
      opening half-angle $\delta_{\rm c}\meq0.2$ and at various inclinations
      $\alpha$ of the cone axis to the plane of the sky.}
    \label{bicone_spec1}
  \end{center}
\end{figure}
\begin{figure}
  \begin{center}
    \makebox[5.222in][l]{
  \vbox to 4.750in{
    \vfill
    \includegraphics{fig08.ps}
  }
  \vspace{-\baselineskip}
}

    \caption{\sf Integrated spectra of intensity (solid line) and degree of
      polarization (dashed line) for a scattering biconical wind with cone
      opening half-angle $\delta_{\rm c}\meq1.0$ and at various inclinations
      $\alpha$ of the cone axis to the plane of the sky.}
    \label{bicone_spec2}
  \end{center}
\end{figure}
In Figures~\ref{bicone_spec1} and~\ref{bicone_spec2},
 examples of these line shapes are given
for narrow ($\delta_{\rm c}\meq0.2$ radians) and wide ($\delta_{\rm
c}\meq1.0$ radians) cones.

\paragraph{Outflowing Disk:}
If the wind is confined to a thin equatorial disk, of angular
half-thickness $\delta_{\rm d}$ then the scattered line profiles can
be calculated by subtracting the result for a biconical wind from
that of a spherical one so that $\delta_{\rm
d}\meq\pi/2\mmin\delta_{\rm c}$. Neglecting
terms higher than quadratic in $\delta_{\rm d}$ (thin disk) then gives
\begin{subeqnarray}
\slabel{integvel_idisk}
I_v &\!\!\!\!\!\meq\!\!\!\!\!& \frac{3}{8}\tau \delta_{\rm d} I_0
\frac{v^2\mmin 2v\mplus 2}{(2v \mmin v^2 \mmin \sin\sq\alpha)^{1/2}}\strut\\
\slabel{integvel_qdisk}
Q_v &\!\!\!\!\!\meq\!\!\!\!\!& -\frac{3}{8}\tau \delta_{\rm d} I_0
\frac{(1\mplus\sin\sq\alpha)(2v\mmin v^2)\mmin 2\sin\sq\alpha}
{\cos\sq\alpha(2v\mmin v^2)(2v \mmin v^2 \mmin \sin\sq\alpha)^{1/2}}\strut,\;\;
\end{subeqnarray}
where $\alpha$ is now the angle between the normal to the disk and the
plane of the sky. Examples of these line shapes are given in
Figure~\ref{disk_spec}.
\begin{figure}
  \begin{center}
    \makebox[5.402in][l]{
  \vbox to 4.750in{
    \vfill
    \includegraphics{fig09.ps}
  }
  \vspace{-\baselineskip}
}

    \caption{\sf Integrated spectra of intensity (solid line) and degree of
      polarization (dashed line) for a scattering wind in the form of
      a thin disk at various angles $\alpha$ between the normal to the
      disk and the plane of the sky. Note that the intensity diverges
      at $v\meq 1\pm \cos\alpha$ and so the values at these points are
      a function of the velocity resolution used in the plots.}
    \label{disk_spec}
  \end{center}
\end{figure}

\subsubsection{Free-falling Inflow}
\label{freefall}
If a dusty flow is assumed to fall radially inwards from rest at
infinity towards a gravitating body of mass $M_{\star}$ then,
neglecting any deceleration, the infall
velocity will be given by $u_{\rm in}\meq(R_0/R)^{1/2}u_0$ where $u_0
\meq (2 G M_{\star}/R_0)^{1/2}$ is the
velocity at the inner cut off radius $R_0$ (which may be identified
with the grain destruction radius). The inflow is supposed to have an
outer radius $R_{\rm c}$ and mass conservation  dictates that
the dust number density have the form $n\meq n_0 (R_0/R)^{3/2}$.
The Doppler frequency shift of the scattered light is
$\Delta\nu\meq-(\nu_0/c)u_{\rm in}(1\mplus \rhat
\cdot \zhat)$, in a similar manner to the outflowing wind but with the
opposite sign. In this case however the dust velocity is not constant,
so the dimensionless velocity shift is dependent on the dust radius:
\def\br{\bar{R}}
$v\meq -(\epsilon_0/\br)^{1/2}(1\mplus\cos\Theta)$, where $\br\meq R/R_{\rm
c}$ and $\epsilon_0 \meq R_0/R_{\rm c}$. This means that the
isovelocity surfaces are no longer cones but have the more complex
\def\ep0{\epsilon_0}
\def\ds{\displaystyle}
\def\half{\frac{\ds 1}{\ds 2}}
\begin{figure}
  \begin{center}
    \makebox[4.152in][l]{
  \vbox to 3.152in{
    \vfill
    \includegraphics{fig10.ps}
  }
  \vspace{-\baselineskip}
}

    \caption{\sf The isovelocity scattering surfaces for a free-falling inflow
      of dust are surfaces of revolution formed by rotating the above curves
      about the line of sight to the center of the flow.
      The curves are marked with the dimensionless Doppler shift induced in
      emission lines scattered by dust lying on them. In this example, the
      ratio of inner to outer dust radius is $\ep0=0.2$. As can be seen, the
      isovelocity surfaces are closed for $v < -2\ep0^{1/2}\msimeq -0.89$. }
    \label{infall_surf}
  \end{center}
\end{figure}
shapes shown in Figure~\ref{infall_surf}. By integrating the scattered
intensity over these isovelocity surfaces, it is possible to determine
the spatially integrated scattered line profile of the infall as
\begin{equation}
  \label{in_eq}
    I_v\meq \frac{3}{8(1\mmin\ep0^{1/2})}
    \times \left\{
    \begin{array}{lc}
      G_1  & \left[\mbox{\bf if} \ \   v <
      -2\ep0^{1/2}\right] \\
      &  \\
      G_2 & \left[ \mbox{\bf
        if}\ \   v \geq -2\ep0^{1/2}\right]
    \end{array}\right. ,
\end{equation}
where
\begin{subeqnarray}
G_1 & \meq & \ln\ep0 \mmin 2v\(\ep0^{-1/2}\mmin1\) \mmin
      \half v^2\(\ep0^{-1}\mmin1\)       \\
G_2 & \meq &  2\ln\(-\half v\) \mplus 2 \mplus 2v \mplus \half v^2 .
\end{subeqnarray}
Note that the scattered line profiles are unpolarized because the
isovelocity surfaces are circularly symmetric from the point of view
of the observer. Example line profiles are shown in
\begin{figure}
  \begin{center}
    \makebox[6.944in][l]{
  \vbox to 2.375in{
    \vfill
    \includegraphics{fig11.ps}
  }
  \vspace{-\baselineskip}
}

    \caption{\sf Spatially integrated scattered line profiles from a
      spherically symmetric free-falling inflow with ratios of inner to
      outer flow radius $\ep0\meq$ $0.05$, $0.25$ and $0.5$.}
    \label{infall_spec}
  \end{center}
\end{figure}
Figure~\ref{infall_spec} for various values of $\ep0$. It can be seen
that for high $\ep0$ (corresponding to a small spread in dust
velocities with radius) the profile is similar to that from the
constant velocity wind but with the sign of the Doppler shift
reversed. As $\ep0$ is decreased there is a greater proportion of
slow-moving dust and so the line profile becomes increasingly skewed
toward $v\meq 0$.
\subsection{Spatially Resolved Scattered Line Profiles:
Position-Velocity Diagrams}
In this section, an analytic approach is used to calculate
the Stokes intensities of the Rayleigh scattered light resolved both
spatially and in velocity for two of the classes of models presented above.

\subsubsection{Moving Source in a Stationary Scattering Cloud}
\label{isovel}
For the purposes of this section, the scattering cloud will be taken
to be spherical and homogeneous. The observer is looking down the
negative $z$-axis (as in Figure~\ref{vel_diag}) and the position of a
scatterer in the cloud is characterized by dimensionless coordinates
$[\bx,\by,\bz]$ where $\bx\meq x/R_{\rm c}$ \etc.
Straightforward geometry
then shows that the dimensionless Doppler shift of light scattered at
$[\bx,\by,\bz]$ is
\begin{equation}
v \meq \frac{\bx \cos\alpha \mplus \bz \sin \alpha} {\(\bx^2\mplus
\by^2 \mplus \bz^2\)^{1/2}}.
\label{vgeneral}
\end{equation}
Hence, solving this equation for $\bz$, a line of sight $[\bx,\by]$
intersects a given isovelocity cone $v$ in zero, one or two places
given by
\begin{equation}
\begin{array}{ll}
\left\{\bz_{\rm a},\bz_{\rm b} \right\} \meq
          \left\{\rule{0cm}{3ex}\right.\!\!\!\! &  \bx\sin\alpha\cos\alpha \pm
          v\left[\bx^2\(1-v^2\)\right.\\
& \left.\left. \mplus\by^2\(\sin\sq\alpha-v^2\)\right]^{1/2}
          \rule{0cm}{3ex}\right\} \mtimes\(v^2-\sin\sq\alpha\)^{-1}
\end{array}
\label{zdef}
\end{equation}
so long as $v$ satisfies
\begin{equation}
              \bx \cos\alpha \pm \left[1\mmin\(\bx^2\mplus\by^2\)\right]^{1/2}
              < |v| < \(1\mmin \frac{\displaystyle\by^2\cos\sq\alpha}
              {\displaystyle\bx^2\mplus\by^2}\)^{1/2},
\label{vlimits}
\end{equation}
where the positive sign in the equation and inequalities
applies to $\bz_{\rm a}$ and
the negative sign to $\bz_{\rm b}$. The left-hand side of
equation~(\ref{vlimits}) reflects the fact that most lines of sight will
not intersect the source velocity vector, while the right-hand side is
the result of the finite size of the scattering cloud.
The scattered line profile along a line of sight can then be
calculated as
\begin{equation}
{\bf I}_v \meq \left| \frac{\dd {\bf I}\,}{\dd v}\right| \meq
\frac{\tau F_0}{4\pi\thc^2}
\left( \sum_{i = a,b}
 \frac{{\bf X}(\bx,\by,\bz_i)}
{\bx^2\mplus\by^2\mplus\bz_i^2}
\left|\frac{\dd \bz_i}{\dd v}\right| \right)
\label{basic_int}
\end{equation}
where $F_0$ is the {\em flux} of the source and $\thc$ is the apparent
angular radius of the scattering cloud. The phase function in
Cartesian coordinates can be written as
\begin{equation}
{\bf X}(\bx,\by,\bz)\meq\frac{3}{4(\bx^2+\by^2+\bz^2)}
\left[ \begin{array}{c}
\bx^2\mplus \by^2 \mplus 2 \bz^2 \\
\bx^2 - \by^2 \\
- 2 \by \bz  \end{array} \right]
\label{cartphase}
\end{equation}
and from equation~(\ref{vgeneral}) it follows that
\begin{equation}
\left|\frac{\dd \bz}{\dd v}\right| \meq
\left| \frac{\(\bx^2\mplus \by^2 \mplus \bz^2\)^{3/2}}
{\(\bx^2\mplus\by^2\)\sin\alpha-\bz\bx\cos\alpha} \right|.
\label{dzdv}
\end{equation}
It is then necessary to eliminate $\bz$ from
 equation~(\ref{basic_int}) using equation~(\ref{zdef}).
For the general case this leads to very complicated expressions which
are best calculated by computer but results for two special cases are
presented here.

\paragraph{Source Velocity in Plane of the Sky ($\alpha\meq 0$):}
\label{an_isovel}
If the dimensionless velocity shift $v$  lies between $\bx$ and
$\bx/\(\bx^2\mplus\by^2\)^{1/2}$, then
\begin{subeqnarray}
I_v&\meq\strut& \frac{3\tau F_0 \left[\bx^2\(2-v^2\)-v^2\by^2
\right]} {8\pi\thc^2\bx\left[\bx^2\(1-v^2\)-v^2\by^2\right]^{1/2}},\\
Q_v&\meq\strut& \frac{3\tau F_0 \,v^2\(\bx^2-\by^2\)} {8\pi\thc^2
\bx\left[\bx^2\(1-v^2\)-v^2\by^2\right]^{1/2}},\\ U_v&\meq\strut&
\frac{-3\tau F_0 \,v^2\bx\by} {4\pi\thc^2
\bx\left[\bx^2\(1-v^2\)-v^2\by^2\right]^{1/2}},
\end{subeqnarray}
otherwise $I_v\meq Q_v \meq U_v \meq 0$.
The degree and angle of polarization are given by
\begin{subeqnarray}
p_v&\meq\strut&
\frac{v^2 \( \bx^2 \mplus \by^2 \)}{\left| 2\bx^2-
v^2\(\bx^2\mplus \by^2\)\right|},\\
\slabel{chi_v}
\chi&\meq& \frac{\pi}{2} \mplus \tan^{-1}\(\frac{\by}{\bx}\)\strut .
\end{subeqnarray}
Note that $\chi$ is independent of $v$, which is true for all values
of $\alpha$.

\paragraph{Source Velocity Along Line of Sight ($\alpha\meq\pi/2$):}
For $\alpha\meq\pi/2$ the scattering geometry has circular symmetry
from the point of view of the observer so it is convenient to
introduce a dimensionless impact parameter
$r\meq(\bx^2\mplus\by^2)^{1/2}$. Then, if $|v|\leq(1\mmin r^2)^{1/2}$,
\begin{subeqnarray}
I_v&\meq\strut& \frac{3\tau F_0 (1\mplus v^2)}
{16\pi\thc^2 r (1\mmin v^2)^{1/2}}, \\
Q_v&\meq\strut& \frac{3\tau F_0 (\bx^2 \mmin \by^2)(1\mmin v^2)^{1/2}}
{16\pi\thc^2 r^3}, \\
U_v&\meq\strut& \frac{-6\tau F_0 \bx \by (1\mmin v^2)^{1/2}}
{16\pi\thc^2 r^3},
\end{subeqnarray}
otherwise $I_v\meq Q_v \meq U_v \meq 0$.
The degree of polarization is
\begin{equation}
p_v\meq \frac{1\mmin v^2}{1\mplus v^2}
\end{equation}
and the angle of polarization is still given by equation~(\ref{chi_v}).
Note that the polarization is independent of position on the cloud for
a given velocity, as is the intensity (apart from a scale factor).
However, the range of allowed velocities {\em is} strongly dependent
on the position on the cloud.
\paragraph{Example Spectra and Position-Velocity Diagrams:}
\begin{figure}
  \begin{center}
    \makebox[5.527in][l]{
  \vbox to 2.680in{
    \vfill
    \includegraphics{fig12.ps}
  }
  \vspace{-\baselineskip}
}

    \caption{\sf The positions, on a spherical cloud, of the synthetic
      apertures used in \S~\protect\ref{isovel}. Two-dimensional
      position-velocity diagrams are presented for Apertures A-C
      (Figures~\protect\ref{w1:ap:a}--\protect\ref{w1:ap:c}) and
      one-dimensional spectra for Apertures D-F
      (Figures~\protect\ref{w1:ap:d}--\protect\ref{w1:ap:f}).}
    \label{slits}
  \end{center}
\end{figure}
In Figure~\ref{slits} the positions on the scattering cloud of six
apertures used in constructing example spectra, three slits and three
point apertures, are shown.  In Figures~\ref{w1:ap:a}
\begin{figure}
  \begin{center}
    \makebox[5.708in][l]{
  \vbox to 6.291in{
    \vfill
    \includegraphics{fig13.ps}
  }
  \vspace{-\baselineskip}
}

    \caption{\sf Doppler- and seeing-broadened position-velocity
      diagrams for a stationary cloud, scattering an emission line
      originating in a moving source.  Aperture~A: centrally placed slit,
      aligned with the source velocity. Dimensionless Doppler width $w\meq0.1$.
      Dimensionless seeing width $a_{\rm s}\meq 0.066$. Intensity contours
      are logarithmic, with successive contours corresponding to a
      ratio of $2^{1/2}$. Value of lowest contour is given on each diagram
      in units of $\tau I_0$ and is $2^{-6}$ times the peak value. Also
      shown are vectors whose lengths are proportional to the degree of
      polarization and whose orientation shows the position angle of the
      polarization.}
    \label{w1:ap:a}
  \end{center}
\end{figure}
\begin{figure}
  \begin{center}
    \makebox[5.708in][l]{
  \vbox to 6.291in{
    \vfill
    \includegraphics{fig14.ps}
  }
  \vspace{-\baselineskip}
}

    \caption{\sf Same as Figure~\protect\ref{w1:ap:a} but for
      Aperture~B: offset slit, parallel to the source velocity.
      Lowest contour is $2^{-4.5}$ times the peak value.}
    \label{w1:ap:b}
  \end{center}
\end{figure}
\begin{figure}
  \begin{center}
    \makebox[5.708in][l]{
  \vbox to 6.291in{
    \vfill
    \includegraphics{fig15.ps}
  }
  \vspace{-\baselineskip}
}

    \caption{\sf Same as Figure~\protect\ref{w1:ap:a} but for
      Aperture~C: offset slit, perpendicular to the source velocity.
      Lowest contour is $2^{-4.5}$ times the peak value.}
    \label{w1:ap:c}
  \end{center}
\end{figure}
\begin{figure}
  \begin{center}
    \makebox[7.486in][l]{
  \vbox to 2.569in{
    \vfill
    \includegraphics{fig16.ps}
  }
  \vspace{-\baselineskip}
}

    \caption{\sf Doppler- and seeing-broadened spectra
      for a stationary cloud, scattering an emission line
      originating in a moving source. Aperture~D: centered on the source.
      Dimensionless Doppler width $w\meq0.1$.
      Dimensionless seeing width $a_{\rm s}\meq 0.066$.}
    \label{w1:ap:d}
  \end{center}
\end{figure}
\begin{figure}
  \begin{center}
    \makebox[7.680in][l]{
  \vbox to 2.569in{
    \vfill
    \includegraphics{fig17.ps}
  }
  \vspace{-\baselineskip}
}

    \caption{\sf Same as Figure~\protect\ref{w1:ap:d} but for Aperture~E:
      centered upstream of the source.}
    \label{w1:ap:e}
  \end{center}
\end{figure}
\begin{figure}
  \begin{center}
    \makebox[7.680in][l]{
  \vbox to 2.569in{
    \vfill
    \includegraphics{fig18.ps}
  }
  \vspace{-\baselineskip}
}

    \caption{\sf Same as Figure~\protect\ref{w1:ap:d} but for Aperture~F:
      centered to the side of the source.}
    \label{w1:ap:f}
  \end{center}
\end{figure}
to~\ref{w1:ap:f}, the spectra and position-velocity diagrams from
these apertures are shown for different values of $\alpha$.  These
spectra include the broadening effects of the frequency profile of the
source emission line and of atmospheric seeing. Both are assumed to be
Gaussian in form; the source profile is taken to have a FWHM of $0.16
u_0$ and the seeing profile to have a FWHM of $0.11\thc$.
\subsubsection{Constant Velocity Scattering Wind}
\label{convel2}
\paragraph{Spherically symmetric wind:}
For a constant velocity wind (assuming a constant dust-gas ratio),
mass conservation requires the dust number density to be of the form
\begin{equation}
n(R) \meq n_0 \(\frac{R_0}{R}\)^2 \meq \frac{n_0 \epsilon_0^2}
{r^2\mplus \bz^2}
\end{equation}
where $n_0$ is the number density at an inner cut-off radius
$R_0\meq\epsilon_0 R_{\rm c}$. The scattering optical depth to the source is
then given by
\begin{equation}
\tau\meq \sigma_0 n_0 R_{\rm c}\epsilon_0(1\mmin \epsilon_0)
\end{equation}
so that the the line profiles are given by
\begin{equation}
{\bf I}_v \meq
\frac{\tau F_0}{4\pi\thc^2}\frac{\epsilon_0}{1\mmin\epsilon_0}
 \frac{{\bf X}(\bx,\by,\bz)}
{(r^2\mplus\bz^2)^2}
\left|\frac{\dd \bz}{\dd v}\right|
\label{basic_int2}.
\end{equation}
Note that in this instance a line of sight can only intersect an
isovelocity cone in, at most, one place and that the situation is very
similar to the $\alpha\meq\pi/2$ case of \S~\ref{isovel}.
Hence, if $(1\mmin(r/\epsilon_0)^2)^{1/2}\leq |1\mmin v| \leq (1\mmin
r^2)^{1/2}$ then
\begin{subeqnarray}\label{wind}
I_v&\meq& \frac{3 \tau F_0 \epsilon_0 (2v\mmin v^2)^{1/2}(v^2\mmin
2v\mplus 2)}{16\pi\thc^2 (1\mmin\epsilon_0) r^3}\\
p_v&\meq&\frac{2v\mmin v^2}{v^2\mmin 2v\mplus 2},
\end{subeqnarray}
otherwise $I_v\meq 0$ ($\chi$ is still given by eq.~[\ref{chi_v}]).
As in the $\alpha\meq\pi/2$ case of
\S~\ref{isovel}, variation in the line profiles with position is
chiefly caused by the conditions for intersection between the line of
sight and isovelocity cone.
\paragraph{Outflowing Bicone:}
With a conical wind, the intensity and polarization profiles will
still be given by equation~(\ref{wind}) but the range of allowed velocities
for the scattered light from a given line of sight is subject to the
additional condition that the line of sight must intersect the
cone of scatterers. To simplify matters, only lines of sight that
intersect the cone symmetry axis ($\by\meq0$) are considered. In this
case, the conditions that there be scattered flux at velocity $v$ from
a line of sight $\bx$ are given in Table~1,
\begin{table*}[htbp]
  \begin{center}
    {TABLE 1}\\
    {\sc Conditions for there to be Scattered Flux from an Outflowing
      Bicone at a Point $\bar{x}$, $v$ in Position-Velocity Space}\\
    \vspace*{5mm}
    \begin{tabular}{c} \hline \\
     \mbox{$ \begin{array}{rcccl}
        \(1\mmin(\bx/\epsilon_0)^2\)^{1/2} & \leq& |1\mmin v| & \leq
        &\(1\mmin \bx^2\) \\ & & & & \\
        \max\left[\cos(\alpha\mplus\delta),0\right] & \leq &\{
        \bx,-\bx \} & \leq & \cos\(\min [\alpha\mmin\delta, 0] \)\\ &
        & & & \\ \left.
        \begin{array}{cc} \sin(\alpha\mmin\delta) &
          [\,\mbox{\bf if}\ \{\bx,-\bx\} > 0\,] \\
          \sin(\alpha\mplus\delta) & [\,\mbox{\bf if}\ \{\bx,-\bx\} < 0
          \,] \strut
        \end{array} \right\} & \leq & \{1\mmin v,\, v\mmin 1\} & \leq
        &\left\{
        \begin{array}{cl} \sin(\alpha\mplus\delta)
          &\parbox{3cm}{$[\,\mbox{\bf if}\
          \cos(\alpha\mplus\delta) > 0 \\ \ \mbox{\bf and} \
          \{\bx,-\bx\} > 0\,]$}\\ 1 & [\,\mbox{\bf otherwise}\,]\strut
        \end{array} \right.
      \end{array} $} \\
      \\
      \hline
    \end{tabular}
  \end{center}
  {\small {\sc Note.}---See text for explanation of the symbols.}
\end{table*}
where all three conditions must be satisfied using either entirely the
left terms in curly brackets or entirely the right. The position
velocity diagrams so obtained (corresponding to a slit placed along
the projected axis of the conical outflow) are illustrated in
\begin{figure}
  \begin{center}
    \makebox[5.708in][l]{
  \vbox to 6.277in{
    \vfill
    \includegraphics{fig19.ps}
  }
  \vspace{-\baselineskip}
}
 \caption{\sf Scattered position
    velocity diagrams from a slit placed along the axis of a narrow
    outflowing bicone of opening half-angle $\delta_{\rm c}\meq 0.2$
    radians at various inclinations $\alpha$ to the plane of the sky.}
    \label{conepv1}
  \end{center}
\end{figure}
\begin{figure}
  \begin{center}
    \makebox[5.708in][l]{
  \vbox to 6.277in{
    \vfill
    \includegraphics{fig20.ps}
  }
  \vspace{-\baselineskip}
}

    \caption{\sf Same as Figure~\protect\ref{conepv1} but for a wide
      outflowing bicone of opening half-angle $\delta_{\rm c}\meq 1.0$
      radians.}
    \label{conepv2}
  \end{center}
\end{figure}
Figures~\ref{conepv1} and~\ref{conepv2} for both a narrow cone and a wide
cone at various inclinations to the plane of the sky.
\paragraph{Outflowing Disk:}
Since the disk is assumed thin, the spatially resolved scattered line
profile at each position on the disk will just be a single spike.
Hence, the position-velocity diagram for a slit placed along the
projected minor axis of the disk ($\by\meq 0$) will be a ridge at $v\meq 1
\mmin \cos\alpha$ for negative $\bx$ and a ridge at $v \meq 1 \mplus
\cos \alpha$ for positive $\bx$, both with a degree of polarization
$\sin\sq\alpha/(1\mplus\cos\sq\alpha)$, whereas for a slit
along the projected major axis of the disk ($\bx\meq 0$) the
position-velocity diagram is merely a ridge at $v\meq 0$ with 100\%
polarization.

\section{Discussion}
The calculations presented in this paper have been selected because
they are amenable to analytic treatment and hence represent extreme
idealizations of situations likely to be encountered in astrophysical
objects. Nevertheless, they demonstrate in a simple fashion the type
of scattered line shapes that will be produced in different
situations.
\subsection{Moving Source}
The moving source model (\S\S~\ref{mover} and~\ref{isovel}) is
applicable to any case in which a line-emitting plasma is moving with
respect to a dusty environment, such as Herbig-Haro objects. This is
discussed in much greater detail in \scite{Hen93b} and \scite{Hen93c},
Papers~{II} and~{III} of this series. It may also furnish an
alternative explanation to that proffered in \scite{Sol92} for the
enormous line widths observed in knots of the collimated outflow from
R~Aquarii.

 It is found that the spatially integrated line shapes from these
models show scattered wings extending a distance equal to the source
velocity $u_0$ to both sides of the rest frequency of the line
(Figure~\ref{intspec}). These wings are polarized to a degree
dependent on the inclination of the source velocity to the line of
sight, being highest when the source is moving in the plane of the sky
and zero if the source is moving directly toward or away from the
observer. Of course, the Doppler shift of the {\em direct\/} light
from the source means that a source moving toward the observer would
be seen to have an unpolarized red wing of width $2u_0$ and a source
moving away from the observer would have a similar unpolarized blue
wing, while a source moving in the plane of the sky would have
polarized red and blue wings.  In this latter case, the polarization
is highest at the extremes of the wings, drops to zero and then rises
again toward line center (although in the center dilution by the
intrinsic light will reduce the observed polarization
substantially). Departure from spherical symmetry of the scattering
cloud (Figures~\ref{prol_vel} and~\ref{ob_vel}) causes the blue- and
red-shifted scattered light to become more (less) intense than the
unshifted light if the cloud is prolate (oblate).

Of the results for spatially resolved line shapes that are presented in
\S~\ref{isovel}, it is perhaps those for Aperture~A of
Figure~\ref{slits} (corresponding to a narrow slit placed along a
diameter of the cloud, parallel to the direction of motion of the
source) that are the most interesting (Figure~\ref{w1:ap:a}). For the
source velocity in the plane of the sky, these show a characteristic
``double triangle'' morphology to the position-velocity diagram of the
scattered light. The scattered light is blue-shifted to the right
(upstream) of the source and red-shifted to the left (downstream). The
polarization is highest for the scattered light that undergoes the
largest Doppler shift (red or blue) since it is the dust directly in
front of or behind the source that scatters light through
$90^{\circ}$, leading to maximum polarization (Eq.~[\ref{pray}]). As
the angle $\alpha$ between the source direction and the plane of the
sky increases, some upstream dust appears to the left of the source
and some downstream dust to the right. Eventually, when the source is
moving directly toward or away from the observer, this leads to the
situation illustrated in the bottom-right panel of
Figure~\ref{w1:ap:a} in which the position-velocity diagram is
symmetrical about $x\meq0$. In this latter case, it is the scattered
light which undergoes no Doppler shift which has the highest
polarization since the dust in the plane of the sky is now directly to
the sides of the motion of the source. Note that, for both
$\alpha\meq0$ and $\alpha\meq\pi/2$, the area of the position-velocity
diagram in which the direct light from the source will lie ($x\meq0$,
$v\meq0$ for $\alpha\meq0$, $v\meq\pm 1$ for $\alpha\meq\pm\pi/2$) is
well away from the area of highest polarization so there will be
little dilution of the scattered light there. Note also that,
unlike in the spatially integrated case, the scattered light is
polarized for all inclinations of the source velocity.

\subsection{Scattering Wind}
These models (\S\S~\ref{convel1} and~\ref{convel2}) are applicable to
any case in which chromospheric emission lines are scattered from a
stellar wind. This may be dust scattering, as is possible in evolved
late-type stars such as Mira variables \cite{Ana93,Lef92}, or electron
scattering, \eg\ in Be stars \cite{Woo93}, B[e] supergiants
\cite{Boy91}, or Wolf-Rayet stars \cite{Sch92}. They are also relevant
to the scattering by dust and electrons in outflows from active
galactic nuclei \cite{Mil90a}.  Of course, for electron scattering the
thermal broadening of the electron velocity distribution will
significantly modify the results presented here for all but the
highest Mach number winds (the thermal speed of the electrons will
exceed the bulk wind velocity so long as the Mach number $M \leq 2
(m_{\rm H}/m_{\rm e})^{1/2} \msim 80$), a fact that is ignored in
\scite{Woo93}. The extension of these results to Thomson scattering
from thermal electrons will be presented in a forthcoming paper.

One obvious result is that some asymmetry of the wind is necessary for
the integrated scattered line profile to be polarized. For a spherical
wind (Figure~\ref{sp_wind_line}), the scattering merely produces an
unpolarized red wing, extending up to twice the wind velocity $u_{\rm
w}$.  If the wind is concentrated toward the polar directions
(Figures~\ref{bicone_spec1} and~\ref{bicone_spec2}), then, for
outflows in the plane of the sky, one finds a polarized red bump at a
redshift of $u_{\rm w}$ whose width depends on the degree of
collimation of the wind. As the outflow axis is rotated toward the
observer (increasing $\alpha$), the bump splits into two, which move
apart to redshifts of zero and $2u_{\rm w}$ as
$\alpha\rightarrow\pi/2$ and whose polarizations diminish to zero.
For equatorially enhanced winds (approximated as a disk;
Figure~\ref{disk_spec}), the intensity profiles of the scattered lines
for a given inclination of the symmetry axis $\alpha$ can be seen to
be qualitatively similar to the profiles for a narrow cone
(Figure~\ref{bicone_spec1}) with inclination $\pi/2\mmin\alpha$, as
would be expected. The polarization profiles do not follow this
pattern, however, and show a maximum at a red shift of $u_{\rm w}$ and
a general decrease in polarization as the disk changes from an edge-on
to a face-on orientation.

The spatially resolved position-velocity diagrams are not illustrated
for the spherically symmetric wind but they are very similar to those
for the moving source model with $\alpha\meq\pi/2$ (bottom-right
panels of Figures~\ref{w1:ap:a}--\ref{w1:ap:c}), except for three
differences: \begin{enumerate}
\item The Doppler shifts will range from 0 to 2$u_{\rm w}$ instead of
from $-u_{\rm w}$ to $u_{\rm w}$.
\item In the spectrogram from Aperture~A, there will be an elliptical
``hole'' in the center because of the inner cut-off to the density
distribution.
\item The scattered brightness will fall off more rapidly with
distance from the center of the cloud because of the inverse-square
density distribution in the wind.
\end{enumerate}
With the conical wind (Figures~\ref{conepv1} and~\ref{conepv2}),
the position-velocity diagrams are the same as for the spherical wind
but with certain portions masked out. The sizes and positions of the
non-empty regions of position-velocity space depend on the opening
angle and orientation of the cones.
\subsection{Scattering Inflow}
The free-falling inflow model (\S~\ref{freefall}) is applicable to the
scattering haloes of young stellar objects, which often show density
profiles indicative of an infalling envelope \cite{Wei92}.

The spatially integrated line shapes (Figure~\ref{infall_spec}) can be
seen to depend sensitively on the value assumed for the ratio of inner
to outer dust radius $\ep0$ (see discussion after eq.~[\ref{in_eq}]).
It should be remembered that the dimensionless Doppler shift $v$ is
scaled to the infall velocity at the inner cut-off radius, which is
proportional to $\ep0^{-1/2}$, so that the velocity scales of
Figure~\ref{infall_spec} are different if all the inflows are assumed
to have the same outer radius and central mass.

Calculations of the spatially resolved line profiles from this model
are not presented here because they are too involved for the simple
analytic approach used in this paper. In addition, it is likely that
many young stellar objects have optically thick haloes \cite{Whi93}
and in such cases the single scattering model presented here is not
strictly applicable. These issues will be addressed in Paper~{III}, in
which Monte Carlo simulations of multiple scattering will be made and
where the effects of a non-zero angular momentum for the infall will
be treated.

\section{Summary}
This paper has presented a small collection of line profiles (both
intensity and polarization) that result from the optically thin
Rayleigh scattering of an emission line by dust in its environment.
The calculations have all been performed analytically and this has
necessarily restricted the complexity of the models employed.
Nonetheless, they form a basis for understanding the ways in which the
geometric and kinematic relations between the line source and the
scatterers determine the scattered line profiles.

Although, the calculations have been performed for the optically thin
case, for one of the kinematic models considered (moving source in a
stationary cloud) large optical depths will make very little
difference to the scattered line profiles. This is because there is no
relative motion between the scatterers themselves and hence the only
Doppler shift is that induced by the first scattering. Of course, for
the wind and infall models, the optically thick case will be quite
different, with much more extended wings to the profiles due to
multiple scattering. However, the polarization will also be reduced
compared with the single scattering case so that the optically thin
models are perhaps more likely to be relevant to sources with
observable polarization changes across their line profiles.

One effect that has been ignored in this paper and that can have a
significant effect on the line profiles is that of a
non-Rayleigh scattering phase function for the dust. This is probably
important at optical wavelengths, where the scattering phase function
can be quite forward-peaked unless the dust grains are smaller than
usual. This issue will considered in depth in Paper~III.

I gratefully acknowledge financial support from SERC, UK and
CONACyT, M\'{e}xico and I would like to thank D. J. Axon and A. C. Raga
for many helpful discussions during the course of this work.

\end{document}